\documentclass[submission, Phys]{SciPost}

\usepackage{amsmath,amsfonts,amsthm,amscd,amssymb}

\DeclareMathOperator{\sech}{sech}

\newcommand{\R}{\mathbb{R}}

\newcommand{\C}{\mathbb{C}}

\newcommand{\CP}{\mathbb{C}P}

\def\bn{\boldsymbol{n}}

\newcommand{\tr}{{\rm tr}}

\def\bee{\begin{equation}}
\def\eee{\end{equation}}

\begin{document}
\parskip 6pt
\parindent 0pt

\begin{center}{\Large \textbf{Gauged Sigma  Models and Magnetic Skyrmions
}}\end{center}

% TODO: write the author list here. Use initials + surname format.
% Separate subsequent authors by a comma, omit comma at the end of the list.
% Mark the corresponding author with a superscript *.
\begin{center}
B. J. Schroers
\end{center}

% TODO: write all affiliations here.
% Format: institute, city, country
\begin{center}
Maxwell Institute for Mathematical Sciences and
Department of Mathematics,
\\Heriot-Watt University,
Edinburgh EH14 4AS, UK. \\
 {\tt b.j.schroers@hw.ac.uk} 
\end{center}

\begin{center}
9 September 2019 
\end{center}

% For convenience during refereeing: line numbers
%\linenumbers

\section*{Abstract}
{\bf We define  a gauged non-linear sigma model for a 2-sphere valued field and a $SU(2)$ connection on an arbitrary Riemann surface whose energy functional reduces to that for  critically coupled magnetic skyrmions  in the plane, with arbitrary Dzyaloshinskii-Moriya interaction, for a  suitably chosen gauge field.   We use the interplay of unitary and holomorphic structures  to derive  a general solution of  the first order Bogomol'nyi equation of the model for any given connection. We illustrate this formula with examples, and also point out applications to the study of impurities.}

\vspace{10pt}
\noindent\rule{\textwidth}{1pt}
\tableofcontents\thispagestyle{fancy}
\noindent\rule{\textwidth}{1pt}
\vspace{10pt}

\section{Introduction}

This paper has two goals. The first is  to  show that a large class of models for static magnetic skyrmions, which includes the most general form of the  Dzyaloshinskii-Moriya   (DM) interaction  \cite{Dzyaloshinskii,Moriya}, can be formulated as a  (non-abelian)  gauged non-linear  sigma model and solved explicitly. The second is to study the relevant  gauged sigma models more generally and to explain the interesting differential geometry which lies behind their solvability.  

Magnetic skyrmions \cite{BY} represent  the most recent and  perhaps  the richest application of the concept of  topological solitons to condensed matter  physics and, potentially, the technology of future magnetic  information storage \cite{FCS}. There is a vast and rapidly growing literature on the subject, partly reviewed in \cite{NT} and, with a particular emphasis on  DM terms in the notation which we will adopt, in \cite{genDMI}. However,  it was only noticed recently in \cite{BSRS} that, among the large class of mathematical models for magnetic skyrmions, there are some, called critically coupled  in \cite{BSRS},  where an infinite family of skyrmion configurations can be found exactly. These configurations satisfy so-called Bogomol'nyi equations and may be viewed as generalisations of  the Belavin-Polyakov self-dual solitons \cite{BP}. 

Our discussion here is an extension and generalisation of   the  paper \cite{BSRS},  both in terms of the range of  applications and  in terms of the underlying mathematics. Instead of  the two-parameter family of DM terms considered in \cite{BSRS} we treat the most general DM term and point out an application to the study of impurities. At the same time,  we clarify the underlying mathematics  by defining  the relevant gauged non-linear sigma model in its most general, geometrically natural form. This leads to a model which we believe to be of independent interest.  It involves  a $S^2$-valued field defined on an arbitrary Riemann surface and   coupled to  a fixed, non-abelian connection 

There is a large literature on gauged non-linear sigma models, but as far as we know the model and energy functonal  we study here has not been considered before. The solutions of our model obey Bogomol'nyi equations which are similar to those occurring in the  abelian non-linear sigma model  of  \cite{Schroers} and   its generalisations (see e.g. the book  \cite{Yang}) and also in  the non-abelian model considered by Nardelli in \cite{Nardelli}. However, our  energy functional is different,  and our  non-abelian gauge field  plays a different role, namely that of an arbitrary but fixed background.

Magnetic skyrmions are  mathematically described as topological solitons in the magnetisation  field $\bn$ of magnetic materials. The latter is a map from a surface $\Sigma$ to the 2-sphere $S^2$. The energy expression for the magnetisation field $\bn$ typically involves a Heisenberg or Dirichlet term (quadratic in derivatives) the crucial  DM term (linear in derivatives)   and various potential terms (no derivatives).  To be specific,  consider  the most general form of such a model  in the plane  $\Sigma=\R^2$. The energy functional is
\bee
\label{Skyrme1}
E[\bn]=\int_{\R^2} \frac{1}{2}(\nabla \bn)^2 + \sum_{a=1}^3\sum_{i=1}^2 {\mathcal D}_{ai}[\partial_i  \bn  \times  \bn)_a +  V(\bn)  \; d x_1d x_2,
\eee
where $V$ is a potential which includes a Zeeman term and anisotropy terms, and $\mathcal D$ is the  tensor 
parametrising the DM interaction. It is sometimes called 
spiralization tensor, see for example  the paper   \cite{genDMI}, to which we also refer the reader  for a discussion of the associated physics and examples.   In order to analyse such a model, it is useful to clarify the mathematical structures which enter its definition.

In two dimensions, the Dirichlet energy expression only depends on a conformal structure of the domain  $\Sigma$. However, the potential terms require an integration measure, so that one would expect  the full model to depend on a metric structure on $\Sigma$. In any case, the energy expression makes use of the (standard)  metric on the target 2-sphere. 

The reformulation of the model \eqref{Skyrme1} in \cite{BSRS}  for a particular  potential and DM term as a gauged non-linear sigma model
  only  requires a  conformal structure  on $\Sigma$ and a choice of an $SU(2)$ connection. However, its solution in terms of holomorphic data makes essential use of the complex structure on the target, i.e. of  the identification of the 2-sphere with the complex projective line $\CP^1$. This suggests a more general formulation of the gauged non-linear sigma model on a Riemann surface $\Sigma$ and also a more general understanding of its solution in terms of the interplay between the metric and complex structure on the target 2-sphere. 

In this paper, we give such a formulation and show that  the resulting model can always, at least locally, be solved in terms of an $SL(2,\C)$-valued map which relates  holomorphic and unitary gauges  for a $\CP^1$-bundle over $\Sigma$. A similar interplay between unitary and holomorphic structures is much studied in the context of self-dual connections on 4-manifolds \cite{Atiyah, DK}, but the 2-dimensional version which we need here is the simplest example of a general theory of connections, often called Chern connections, which are  compatible with unitary and holomorphic structures on a complex vector bundle, see  \cite{Moroianu}.

Our presentation is organised as follows. We begin, in Sect.~2, with a definition and discussion of the gauged non-linear sigma model in the most general setting of an arbitrary Riemann surface $\Sigma$. We define the energy functional and derive a lower bound for it in terms of a topological invariant and the  boundary behaviour of the magnetisation field (provided $\Sigma$ has a boundary). We then derive a first order Bogomol'nyi equation  for configurations which saturate this bound. In Sect.~3 we begin with a brief review of relevant complex geometry, in particular the concept of a Chern connection, and then derive  a general formula for solutions of the Bogomol'nyi equation for any given $SU(2)$ connection. In Sect.~4 we apply the results of Sects.~2 and 3 to magnetic skyrmions by writing the energy functional \eqref{Skyrme1} for a particular choice of potential (associated to the spiralization tensor $\mathcal D$) as a gauged non-linear sigma model. This requires  a translation of  the spiralization tensor $\mathcal D$  into  a $SU(2)$  connection, which we shall explain. We 
 use the general formula of Sect.~3 to obtain infinite families of exact  magnetic skyrmions for any given spiralization tensor and an associated potential, and discuss some examples. We also explain how our formalism can be used to define and solve models of topological solitons interacting with impurities.   Our final Sect.~5 contains a conclusion and an outlook. 

The logic of our mathematical derivation requires that we begin our discussion with gauged non-linear sigma models, but readers who are predominantly interested in magnetic skyrmions and willing to take the solution formula on trust are invited to skip to Sect.~4. 

\section{Gauged sigma models on a Riemann surface}
\label{sigmadef}
\subsection{Conventions}
\label{consec}
The gauged sigma model  we want to consider can be defined on any Riemann surface $ \Sigma$, i.e. on any one-dimensional complex manifold,   with our without boundary. We will define the model  using invariant notation, but for concrete and  explicit expressions we use a local complex coordinate  $z=x_1+ix_2$. We also use the standard notation
\bee
dz= dx_1+idx_2, \qquad \partial_z = \frac 12 (\partial_1 - i \partial_2), \quad  \partial_{\bar z}  = \frac 12 (\partial_1 +i \partial_2).
\eee 
The  Hodge $\star$ operator on 1-forms is determined by the complex structure; in local coordinates it is 
\bee 
\label{starsigns} 
\star dz= dx_2-idx_1= -i dz, \qquad \star d \bar z = dx_2+idx_1 = id \bar z.
\eee
Note  also that, for any two 1-forms $\alpha, \beta$ on $\Sigma$,
\bee
\label{starprop}
\star \star \alpha = - \alpha \quad \text{and} \quad 
\alpha \wedge \star \beta =  \beta \wedge \star \alpha.
\eee

Consider now a principal  $SU(2)$-bundle  $P$ over $\Sigma$  with a connection, and the  associated adjoint bundle as well as the unit 2-sphere bundle $P\times^{\text{Ad}}S^2$  in the adjoint bundle. We think of the fibre $S^2$ as the round 2-sphere of radius 1 inside the Lie algebra $su(2)$, with  $SU(2)$ acting in the adjoint representation.  Locally, in an open set $U\subset \Sigma$,  a section of $P\times^{\text{Ad}}S^2$   is a map
\bee
n: U  \rightarrow S^2 \subset su(2),
\eee 
and the connection is given  by a $su(2)$-valued 1-form $A$ on $U$. The  exterior  covariant derivative of $n$ and the curvature is given by the usual expressions  
\bee
\label{covcur}
Dn  = dn +[A,n], \quad F= dA + \frac 12 [A,A]. 
\eee
Here and in the rest of the paper  we suppress the wedge product in the commutator of Lie  algebra-valued 1-forms. 
Gauge transformations are determined locally by functions $u:U \rightarrow SU(2)$ and take the form 
\bee
\label{gauge}
n\rightarrow unu^{-1}, \qquad A \rightarrow uAu^{-1} + u du^{-1}.
\eee 

In local coordinates on $\Sigma$, we expand 
\bee
\label{Asplit} 
A= A_1 dx_1 + A_2 dx_2 = A_z dz+ A_{\bar z} d\bar z,
\eee
where we defined $A_z=\frac12 (A_1-iA_2)$, $A_{\bar z}=\frac 12 (A_1+iA_2)$,
and similarly for the curvature
\bee
F= F_{12} \, dx_1\wedge dx_2, \qquad F_{12 } = \partial_1A_2 -  \partial_2 A_1 + [A_1,A_2].
%= F_{z\bar z}dz \wedge d\bar z,
\eee
We  use $su(2)$ generators $t_a=-\frac{i}{2}\tau_a$ , $a=1,2,3$, where $\tau_a$ are the Pauli matrices, which are normalised so that 
$[t_1,t_2]=t_3$ + cycl., and also  define raising and lowering  operators  
\bee
\label{Liegen}
t_+=t_1 +i t_2, \quad t_-=t_1-it_2.
\eee
These  naturally lie in the complexified $su(2)$ Lie algebra, so in $sl(2,\C)$.
For the inner product  of $m,n \in su(2)$ we use a re-scaled trace and write
\bee
\label{scapro}
(m,n) =-2 \tr (mn),
\eee 
as well as $|n|^2= (n,n)$. Our normalisation is such that  $
(t_a,t_b)= \delta_{ab}$.

\subsection{Energy and variational equations}
The energy functional  defining  the gauged sigma model  we study in this paper  is  
\bee
\label{energy}
E[A,n]=  \frac 12\int_\Sigma (D n, \wedge  \star D n) - \int_\Sigma ( F,  n),
\eee
or, in  local coordinates and in terms of $D_i= \partial_i + [A_i,\cdot ]$, $i=1,2$, 
\bee
\label{localenergy}
E[A,n]=  \int_\Sigma \left( \frac 12 |D_1n|^2 +\frac 12  |D_2n|^2   - ( F_{12}, n) \right) dx_1 \wedge dx_2 .
\eee
Clearly, the covariant Dirichlet term depends on the complex structure (through  $\star$), but the curvature term does not and is topological in that sense.  The energy is manifestly invariant  under  the gauge transformations \eqref{gauge}.

As our notation indicates, we think of the energy as a functional of the connection $A$ and the section $n$. Postponing a discussion of boundary terms we initially assume that the Riemann surface $\Sigma$ has no boundary. 
Then variation with respect to the connection gives 
\bee
\label{deltaA}
\delta E = \int_\Sigma ( Dn, \wedge \star [\delta A,n ] ) - (n, (d \delta A + [\delta A, A]) ) =\int_{\Sigma} (\delta A, \wedge \star [n,Dn]) -(\delta A,\wedge  Dn).
\eee
Setting this to zero for all $\delta A$ gives the Euler-Lagrange equation 
\bee
\label{Bogon}
 \star [n,Dn] = Dn. 
\eee
Using \eqref{starsigns}, this can also be written as 
\bee
D_{\bar z} n = i[n,D_{\bar z} n] \Leftrightarrow  D_{z} n = -i[n,D_{ z} n].
\eee
Since  $[n,\cdot]$  is the complex structure  in the cotangent space to $S^2$ at $n$,  the equation \eqref{Bogon}  is a holomorphicity condition.  This will become more obvious and useful when we study this equation   in complex coordinates for $S^2$  in Sect.~\ref{stereo}.

The variation with respect to $n$,  using $\delta n =[\epsilon ,  n] $ to preserve $|n| =1$  and neglecting boundary terms  gives
\bee
\label{deltan}
\delta E= \int_\Sigma ( D\delta n, \wedge \star Dn)  - (\delta n,  F)  = -\int_\Sigma ( \delta n,(  D \star Dn  + F)) =- \int_\Sigma (\epsilon,  [n, D \star Dn  + F]).
\eee
Setting this to zero for all $\epsilon $ leads to the variational equation
\bee
\label{Laplace}
[n, D \star Dn  +F]=0. 
\eee

It is not difficult to check that the first order equation \eqref{Bogon} actually implies the second order equation \eqref{Laplace}. Applying $\star$ to \eqref{Bogon} and differentiating, we obtain
\bee
D \star Dn = -D[n,Dn] = -[Dn,  Dn] -[n,D^2n] = - [Dn, Dn] - [n,[F,n]].
\eee
Now we use that   $[Dn, Dn]$ is in the direction of $n$ to deduce
\bee
[n,D \star Dn]=-[n,[n[F,n]]]= -[n,F],
\eee
as claimed.  The equation \eqref{Bogon} is therefore the only equation we need to consider. We now show that it can also  be interpreted as a Bogomol'nyi equation in this model.

\subsection{The Bogomol'nyi equation}
The logical dependence of the two variational equations can be understood better by noting that \eqref{Bogon} can also be obtained via a Bogomol'nyi argument. To show this, we need 
 't Hooft's identity \cite{tHooft} relating the integrand for the degree,
 \bee
4\pi \text{deg}[n]= \int_\Sigma  \frac 12 ( n,  [dn ,   dn]),
 \eee
to its covariant version:
\bee
\label{tHooftid}
\frac 1 2 (  n,  [Dn ,Dn] )  = \frac 12  ( n , [dn, dn])  + ( F,  n)  - d( A ,  n).
\eee
Now use \eqref{starprop} and the cyclical property of the triple product to note 
\bee
( (Dn - \star [n,Dn]), \wedge \star (Dn - \star [n,Dn])) = 2 ( Dn, \wedge \star Dn)- 2 (n, [Dn,Dn]).
\eee
This allows us to write the energy as 
\bee
E[A,n]=\frac 1 4 \int_\Sigma ((Dn - \star [n,Dn]),\wedge \star (Dn - \star [n,Dn]))  + \frac 12  \int_\Sigma
 ( n, [Dn, Dn])  - \int_\Sigma (F, n) .
 \eee
 Combining this with the identity \eqref{tHooftid}, we
 deduce 
 \bee
 \label{Bogotrick}
 E[A,n]= \frac 1 4 \int_\Sigma ((Dn - \star [n,Dn]),\wedge \star (Dn - \star [n,Dn]))  + \frac 12  \int_\Sigma
  ( n, [dn, dn]) - \int_{\partial \Sigma}( A, n),
\eee
with the last term of course vanishing when $\Sigma$ has no boundary. We conclude that the energy is bounded below by terms which only depend on topology (the degree of $n$) or on boundary behaviour (if there is a boundary). If both are kept  fixed, the energy is minimised iff the first order equation \eqref{Bogon} holds. The energy of such  Bogomol'nyi configurations is 
\bee
\label{Bogenergy}
E_B[A,n] = \frac 12  \int_\Sigma
  ( n, [dn ,dn]) - \int_{\partial \Sigma} ( A, n).
\eee

The equation \eqref{Bogon} is thus seen to be a Bogomol'nyi equation   in the general sense of characterising minima of energy functionals subject to topological or boundary conditions \cite{MS}. Such equations  usually imply the  variational equations. This is the case here, too, but in a somewhat unusual way. The Bogomol'nyi equation of the model {\em is} the variational equation \eqref{Bogon} with respect to the connection, and implies the second order variational equation with respect to the field $n$, as we already showed.

The Bogomol'nyi equation \eqref{Bogon} clearly does not uniquely determine both the connection and the section $n$. It is easy to write down infinitely many solutions for the connection $A$ for {\em any} given smooth section $n$ in the form 
\bee
\label{Aforn}
A= an -(p[n,dn] + q \star dn) + r(\star  [n,dn] + dn),
\eee
where $a$ is a 1-form and $p,q,r $ are real functions, with $r $ arbitrary but $p,q$  satisfying $p+q=1$.  The choice $A=0$ is possible when $\star[n,dn]= dn$, and is then realised  with $p=q=\frac 12 $, and $ a=0,  r=0$.  This is expected because in that case $n$ satisfies the ungauged Bogomol'nyi equation. 

Note that the  1-form $[n,dn]$ is naturally associated to $n$ as the Levi-Civita connection on the plane bundle orthogonal to $n$ inside the trivial bundle $\Sigma \times su(2)$, and that the remaining terms other than $an$ are obtained from the Levi-Civita 1-form  by applying the complex structures $\star$  on the domain or $[n,\cdot]$  on the target.  Also note that gauge transformations $g=\exp(\alpha n)$ which fix   $n$ (so $\alpha$ is a function) determine gauge equivalent solutions for fixed $n$.

Alternatively, one can use transformations  \eqref{gauge}
to rotate $n$ in a fixed direction (say $t_3$) in the Lie algebra on some open set $U\in \Sigma$. In this gauge, $A$ is determined by the algebraic condition
\bee
\label{constgauge}
\star [t_3,[A,t_3]]= [A,t_3].
\eee

Solving for $A$ when $n$ 
is given may be of interest  in some applications, e.g. when one would like a particular configuration to be  a solution in the presence of an impurity, see our discussion in Sect.~\ref{impuresect}.  However, 
we will now focus on the opposite situation where $A$ is a given background gauge field, and we solve  \eqref{Bogon} for $n$ in this background.

\subsection{Boundary terms}
\label{boundarysect}

So far we have ignored boundary terms which arise in the derivation of the variational  equations for the  functional \eqref{energy}. When $\Sigma $ does have a boundary $\partial \Sigma$,   boundary  terms will generally only vanish if we impose a suitable boundary condition. When $\Sigma$ is an open set - such as $\C$, or the upper half-plane - we need to impose suitable fall-off conditions as we approach `infinity'.  In order to discuss these matters  in any detail we would need to fix the surface $\Sigma$ and the background gauge field $A$ we want to consider. We will not do this here, but make some general observations. 

The boundary term which arises in the variation \eqref{deltan} of \eqref{energy} with respect to $n$ is 
\bee
\label{boundary}
\int_{\partial \Sigma}(\delta n,   \star D n)= \int_{\partial \Sigma}(\epsilon, \star [n, D n]).
\eee
If  $\partial \Sigma$ is an actual boundary and we impose a Dirichlet  boundary condition $n_{|\partial \Sigma}  =n_\infty$ this term vanishes because we must require $\delta n =[\epsilon,n]=0$ on the boundary. However, when  $\Sigma =\C$,  the requirements $\lim_{|z| \rightarrow \infty} n (z) = n_\infty$ and  $\lim_{|z| \rightarrow \infty} \epsilon (z) =0 $ are  not sufficient to ensure the vanishing or even well-definedness of the integral  \eqref{boundary}: when the gauge potential $A$ does not vanish  in the limit $|z| \rightarrow \infty$ the term  in the integrand of \eqref{boundary} containing  $A$ may have a non-zero integral even when $\epsilon$ vanishes in this limit.  The situation can be improved by considering  the modified energy functional
\bee
\label{tildeE}
\tilde E[A,n]=  \int_\Sigma \frac 12 (D n, \wedge \star D n)  -  (F,n)+ \int_{\partial \Sigma} ( A, n). 
\eee
The inclusion of the boundary term means that the modified energy is bounded below by the integral defining the degree, see \eqref{Bogotrick}.   Now variation  with respect to $n$  gives
\bee
\delta \tilde E = - \int_\Sigma (\delta n,  (D \star Dn +F) ) + \int_{\partial \Sigma} (\delta n, (\star D n + A)). 
\eee
Using again $\delta n =[\epsilon, n]$ we only  need to  assume the Bogomol'nyi equation \eqref{Bogon} on the boundary to obtain 
\bee
\delta \tilde E = - \int_\Sigma (\delta n,  (D \star Dn +F) ) + \int_{\partial \Sigma} (\epsilon, dn ). 
\eee
Comparing the new  boundary term with \eqref{boundary} we observe that the modification of the energy expression has removed the troublesome term in \eqref{boundary} involving the gauge potential. 

In summary, the modified energy functional  \eqref{tildeE} is a better starting point for a  well-defined variational problem in the presence of a boundary, though the details will depend on the surface and boundary under consideration.  Requiring the Bogomol'nyi equation in the boundary region and mild fall-off conditions for $\epsilon$ and 
and $dn$ are sufficient to remove boundary terms which arise in the variation. The bulk equations for $n$ are, of course, the equations \eqref{Laplace} already derived in the absence of a boundary.

\section{Solving the Bogomol'nyi equation}

\subsection{Holomorphic versus unitary structures}
\label{holounit}
We will now show how to solve the Bogomol'nyi equation \eqref{Bogon} for a given connection. The idea is to  exploit the interplay between a holomorphic and a unitary structure on a complex vector bundle  over $\Sigma$, and the special properties of  the unique connection, often called the  Chern connection,  which is compatible with both. The underlying  theory  is covered, for example in \cite{Moroianu, DK} and also more informally in  \cite{Atiyah}. 

We  consider only the setting which is relevant for our discussion, so look  at  holomorphic $\C^2$-bundles over $\Sigma$ with a unitary structure (a Hermitian inner product on the fibres).  Any such  vector bundle, denoted $E$ in the following,  has an associated projective bundle; this is a holomorphic $\CP^1$-bundle and, with the unitary structure, will be identified with the  $S^2$-bundle $P\times^{\text{Ad}}S^2$ of Sect.~\ref{consec}.   We use the standard notation of $\partial $ and $\bar \partial$ for the exterior derivative followed by projection onto  differential forms of type $(1,0)$ and $(0,1)$  on $\Sigma$, so in our local coordinates and applied to functions $f$,
\bee
\partial f = \partial_z f dz, \qquad \bar \partial f = \partial_{\bar z}f  d\bar z. 
\eee
Now consider a connection on  the vector bundle $E$.  The associated covariant derivative 
 $D= d+A$ can be split into 
 \bee
 \partial_A = \partial + A_z dz, \qquad \bar\partial_A= \bar \partial + A_{\bar z} d\bar z,
 \eee
where $A= A_z dz +A_{\bar z} d\bar z $ is simply the split \eqref{Asplit} of the matrix-valued 1-form $A$ into forms of type $(1,0)$ and $(0,1)$. Such a connection  is called unitary if it preserves the Hermitian inner product on the fibres, and it  is called compatible  with the holomorphic  structure of $E$ if  $\bar \partial_A  \vec{w} =0$ for every holomorphic section\footnote{Sections of a bundle $E$ over $\Sigma$  are holomorphic if they are holomorphic  as maps   from $\Sigma$ into the total space of the bundle $E$.} $\vec{w}$  of $E$. If a connection is compatible with both structures then, in a unitary gauge (a local choice of an orthonormal basis of the fibre), the gauge potential has to satisfy the anti-Hermiticity condition 
\bee
\label{unitcond}
(A^u_{\bar z} d\bar z)^\dagger = - A^u_{z }dz.
\eee
On the other hand, in a holomorphic gauge (a local choice of a holomorphic basis of the fibre), we have 
\bee
\bar{\partial}_{A^h} =\bar{\partial}.
\eee
It is now straightforward to check that any connection which is compatible with both the unitary and the holomorphic structure must have curvature of type $(1,1)$. This follows by a short computation which is important for us and which we therefore spell out. The gauge change from the holomorphic to the unitary gauge must be via a locally defined map  $g:U\subset \Sigma \rightarrow GL(2,\C)$ satisfying 
\bee
\bar{\partial}_{A^u} = \bar\partial +  g\bar\partial g^{-1}.
\eee
But then the condition \eqref{unitcond} fixes the connection uniquely and implies  an explicit formula for the gauge potential in the unitary gauge, valid in the open set $U\subset \Sigma$: 
\bee
\label{hermgauge}
{A^u} =  g\bar\partial g^{-1}  + (g^{-1})^\dagger \partial  g^\dagger.
\eee
This shows in particular that if $A^u$ is the gauge potential of an $SU(2)$ connection, then $g$ has determinant 1 and  is therefore $SL(2,\C)$-valued. This expression for a gauge potential in two dimensions is also frequently used in the physics literature on planar  $SU(2)$ Yang-Mills theory, see e.g. \cite{KaNa}. 

We can transform back, using $g^{-1}$, from the unitary to the holomorphic gauge to deduce the $(1,0)$ component  and hence the entire  gauge field in the holomorphic gauge as 
\bee
\label{holgauge}
A^h = g^{-1}  (g^{-1})^\dagger \partial  g^\dagger g + g^{-1}\partial g = h^{-1}\partial h,
\eee
where we defined  $h=g^\dagger g$. 
The matrix $h$ is manifestly Hermitian and positive definite, and  defines a Hermitian inner product on the fibre in the holomorphic gauge \cite{DK}. The curvature 2-form then comes out as 
\bee
F= \bar{\partial} (h^{-1} \partial h), 
\eee
which is  manifestly of type $(1,1)$ (and will remain so after gauge transformations), as claimed.

One can also prove the converse result \cite{Atiyah,DK}. If a complex vector bundle $E$ over a complex manifold, with a unitary structure,  has a connection which is unitary  and has curvature of type $(1,1)$ then there is a unique holomorphic structure on $E$  such that the connection has the forms \eqref{hermgauge} and \eqref{holgauge} in the unitary and holomorphic gauge respectively. 

In one complex dimension, any connection has curvature of type $(1,1)$ and it  follows that the unitary connections on $\Sigma$ which we considered in Sect.~2 define  complex structures on the total space of the $S^2$-bundle $P\times^{\text{Ad}}S^2$  over $\Sigma$.  Since we are interested in $SU(2)$ connections, they can always locally be expressed in the form \eqref{hermgauge} for $g:U \rightarrow  SL(2,\C)$. This is the result which we will put to practical use in the next section. As a final preparation we recall  the Iwasawa decomposition of 
 $g\in SL(2,\C)$   via
\bee
g=u\rho,
\eee
with $u\in SU(2)$ and $\rho$ an upper-triangular matrix with unit determinant of the form 
\bee
\label{rhodef}
\rho = \begin{pmatrix} \lambda & c \\ 0  & \frac 1 \lambda \end{pmatrix}, \qquad \lambda\ \in \R^+, \; c \in \C. 
\eee
Since the unitary factor $u$  acts as an overall unitary gauge transformation in \eqref{hermgauge} we can  express any  Hermitian gauge potential on $\Sigma$  up to $SU(2)$ gauge transformation locally as 
\bee
\label{rhogauge}
A =  \rho\bar\partial \rho^{-1}  + (\rho^{-1})^\dagger \partial  \rho^\dagger,
\eee
where $\rho$ is a matrix-valued function of the form \eqref{rhodef}.

\subsection{Holomorphic structure of the gauged sigma model}  
\label{stereo}
In order to apply the theory of the previous section to  the gauged sigma model of Sect.~\ref{sigmadef}, we need a little more notation. 
We write vectors in $\C^2$  as
\bee 
\vec{w}= \begin{pmatrix} w_1 \\ w_2 \end{pmatrix},
\eee 
and use    the standard Hermitian product $\langle \vec{v}, \vec{w} \rangle = v_1\bar{w}_1 + v_2 \bar{w}_2$ on $\C^2$.
The Hopf projection maps the  unit  sphere $S^3$  in $\C^2$ to the unit sphere $S^2$ in the Lie algebra $su(2)$ via
\bee
\label{Hopf}
\pi: S^3\subset \C^2  \rightarrow S^2\subset su(2), \quad \vec{w} \mapsto n= W t_3 W^{-1}, \qquad W = \begin{pmatrix}
w_1 & -\bar{w}_2  \\ w_2 & \phantom{-} \bar{w}_1 
\end{pmatrix},
\eee 
or, with $n=n_1t_1+n_2t_2+n_3t_3$,
\bee
\label{Hopfexplicit}
n_1+in_2=2w_2\bar{w}_1, \qquad n_3= |w_1|^2-|w_2|^2.
\eee
The standard  action  of $u\in SU(2)$  on $\C^2$,
\bee
\label{uleft} 
u: \C^2 \rightarrow \C^2, \quad  \vec{w} \mapsto u\vec{w},
\eee 
 induces the adjoint action, 
\bee
\label{uad}
u:su(2) \rightarrow su(2), \qquad  n\mapsto unu^{-1} = R(u)n,
\eee
 which preserves the inner product  \eqref{scapro}  in $ su(2)$. 
 
 We use the conventions of \cite{BSRS} to define  a stereographic coordinate $w\in \C \cup \{\infty\}$  for the 2-sphere by projection from the south pole,
\bee
\label{wn}
w = \text{St}(n) = \frac{n_1+in_2}{1+n_3},
\eee
with inverse 
\bee 
\label{nw}
 n_1+in_2= \frac{ 2 w }{1+|w|^2}, \quad n_3= \frac{1-|w|^2}{1+|w|^2}.
 \eee
One checks that the Hopf projection \eqref{Hopf} followed by stereographic projection can now also be written as 
\bee
\text{St}  \circ\pi:\vec{w} \mapsto  w=  \frac{w_2}{w_1}.
\eee 
 
An element 
\bee
g= \begin{pmatrix} a & b \\ c& d \end{pmatrix} \in SL(2,\C) 
\eee
acts on $\vec{w} \in \C^2$  by ordinary matrix multiplication
\bee
g:\vec{w} \mapsto g \vec{w},
\eee
 and on our projective coordinate $w$  by fractional linear transformation, which we write as 
\bee
\label{fractrans}
w\mapsto  g[w]:= \frac{c + dw}{a +bw}.
\eee
For $u\in SU(2) \subset SL(2,\C)$, this action agrees with \eqref{uad} when $w$ and $n$ are related via the stereographic map \eqref{wn}. However, non-unitary elements in $SL(2,\C)$ act as  conformal transformations which do not  preserve  the round metric induced by the embedding $S^2 \subset su(2)$.

To write the gauged non-linear sigma model in terms of  the stereographic coordinate $w$,  we note that our $sl(2,\C)$  Lie algebra generators \eqref{Liegen} are explicitly 
 \bee
 \label{tpm}
t_+ =t_1+it_2= \begin{pmatrix}  0  &  -i \\  0 & \phantom{-}0 \end{pmatrix}, \quad 
t_- =t_1-it_2= \begin{pmatrix} \phantom{-} 0  &  0 \\  -i & 0 \end{pmatrix}, \quad t_3 =
\begin{pmatrix}  -\frac i 2   & 0  \\ \phantom{-} 0  & \frac i 2 \end{pmatrix}.
\eee
Writing their action  on the projective coordinate $w$ simply as juxtaposition, we have, for general $t\in sl(2,\C)$, 
\bee
\label{Lieaction}
t w = \left. \frac{d}{d\epsilon} \right|_{\epsilon=0} \exp(\epsilon t)[w], \quad t \in sl(2,\C),
\eee
and compute  
\bee
 t_- w = -i, \quad  t_3 w= iw,  \quad  t_+ w = iw^2.
\eee

Defining the Lie algebra components (as opposed to the 1-form components \eqref{Asplit})  of the gauge potential $A$ via
\bee
\label{Aexp}
A=  \frac 12 ( A_+ t_-+A_-t_+ ) +A_3 t_3, 
\eee
and similarly for the curvature 
\bee
F= \frac 12 (F_+ t_-+ F_-t_+ ) +F_3 t_3,
\eee
we can write the covariant derivative as 
\bee
\label{complexcov}
Dw= dw + Aw =  dw  -\frac i2 A_+ +i A_3 w  + \frac i 2 A_-w^2,
\eee
and have the identity
\bee
(F,n)= \frac{wF_-+ \bar{w} F_+ +F_3 (1-|w|^2)}{1+|w|^2}.
\eee

Using the standard expression for the  Dirichlet  term in terms of stereographic coordinates \cite{MS},
the energy \eqref{energy} of the gauged non-linear  sigma model then takes the form 
\bee
E[A,w]=  \int_\Sigma 2  \frac{ Dw \wedge \star  \overline{Dw} }{(1+|w|^2)^2} - \int_{\Sigma} \frac{wF_-+ \bar{w} F_+ +F_3 (1-|w|^2)}{1+|w|^2},
\eee
and the   identity \eqref{tHooftid} reads
\begin{align}
2i \frac{Dw \wedge \overline{D w}}{(1+|w|^2)^2} &= 2i \frac{dw \wedge \overline{d w}}{(1+|w|^2)^2} + \frac{wF_-+ \bar{w} F_+ +F_3 (1-|w|^2)}{1+|w|^2} \nonumber \\ & \qquad \qquad \qquad \qquad -d \left(\frac{wA_-+ \bar{w} A_+ +A_3 (1-|w|^2)}{1+|w|^2}\right).
\end{align} 
With 
\bee
(Dw -i \star Dw)\wedge \star \overline{(Dw -i \star Dw)} =
 2 Dw\wedge \star \overline{Dw} -2i  Dw\wedge  \overline{Dw},
\eee
the  energy   can be therefore be  written as 
\begin{align}
E[A,w]=& \int_\Sigma \frac{(Dw -i \star Dw)\wedge \star \overline{(Dw -i \star Dw)} }{(1+|w|^2)^2} +2i \int_{\Sigma} \frac{dw \wedge \overline{d w}}{(1+|w|^2)^2} \nonumber \\
&\qquad \qquad \qquad \qquad\qquad   -\int_{\partial \Sigma} \frac{wA_-+ \bar{w} A_+ +A_3 (1-|w|^2)}{1+|w|^2}.
\end{align}
The second term is   $4\pi$ times  the degree of $w$, and the last term is a boundary term. If both degree and boundary behaviour are kept fixed, minima of the energy are therefore determined by the equation
\bee
\label{Bogonhol}
Dw = i\star Dw \Leftrightarrow D_{\bar z} w =0,
\eee
where we used the basic properties \eqref{starsigns} of the $\star$-operator on 1-forms. This is the Bogomol'nyi equation \eqref{Bogon} in stereographic coordinates, as can also be checked by explicitly changing coordinates according to \eqref{nw} in \eqref{Bogon}.

A key feature of the  equation \eqref{Bogonhol}, which was not obvious in the formulation \eqref{Bogon},  is its  gauge invariance under the larger group of $SL(2,\C)$-valued gauge transformation
\bee
\label{holgaugetrafo}
A_{\bar z} \mapsto gA_{\bar z}g^{-1} + g \partial_{\bar z} g^{-1}, \qquad w\mapsto g[w], 
\eee
where $g:U\subset\Sigma\rightarrow SL(2,\C) $,  and we used the notation \eqref{fractrans} for fractional linear transformations.

\subsection{A general solution}
We can now apply the geometrical considerations of Sect.~\ref{holounit} to solve the Bogomol'nyi equation \eqref{Bogonhol} for a given $su(2)$-connection $A$ on the principal bundle $P$ as follows. We consider the  $\C^2$-bundle associated to $P$ via \eqref{uleft}. By the results of Sect.~\ref{holounit}, the  connection $A$ defines a holomorphic structure on this bundle, and hence also on the associated  projective $\CP^1$-bundle. Locally, we can go to a holomorphic gauge via an $SL(2,\C)$ gauge transformation. In this gauge $\bar{\partial}_A = \bar{\partial}$, so that   the Bogomol'nyi equation \eqref{Bogonhol} can easily be solved. 

Explicitly, this means that, for a given unitary connection $A$, we need to find a locally defined map 
\bee
g:U\subset \Sigma \rightarrow SL(2,\C),
\eee
so that the anti-holomorphic component of $A$ is 
\bee
\label{Agrel}
A_{\bar z} = g\partial_{\bar z} g^{-1}.
\eee
Then the Bogomol'nyi equation \eqref{Bogonhol} becomes simply
\bee
\partial_{\bar z} w + g\partial_{\bar z} g^{-1}w = 0.
\eee 
Using again  our notation \eqref{fractrans} for the action of $g$ on $w$ by fractional linear transformations, this means that  $
f = g^{-1}[ w] $ 
is a holomorphic function.  Thus we obtain  the general solution, valid in some open set $U\subset \Sigma$,  as 
\bee
\label{mastersol}
w= g[ f],   \quad \text{with} \quad   f:U  \rightarrow \CP^1 \quad  \text{holomorphic}. 
\eee
The  field $n$ can then be reconstructed via \eqref{nw}. 
We will illustrate this formalism  in the next section by applying it to models of magnetic skyrmions, written as gauged non-linear sigma models. 

\section{Applications to magnetic skyrmions and impurities}
\subsection{Critically coupled magnetic skyrmions  with any  DM term}

In the remainder of the paper we focus on models defined in  the plane  $\Sigma =\C$. For our solution of the gauged sigma model,  both holomorphic and unitary gauges can be chosen globally in this case and therefore we obtain a global solution of the form \eqref{mastersol}.   

In order to apply our method  to magnetic skyrmions we consider the general  model already presented in the Introduction, but now write the energy \eqref{Skyrme1} in Lie algebraic notation, which simply means interpreting the magnetisation vector $\bn$ as a unit length element $n$ of the Lie algebra $su(2)$, and writing vector products as commutators:  
\bee
\label{Skyrme2}
E_S[n]=\int_{\R^2} \frac{1}{2}|\partial_1 n|^2 + \frac 12 |\partial_2 n|^2  + \sum_{a=1}^3\sum_{i=1}^2 {\mathcal D}_{ai}[\partial_i n , n]_a +  V(n)  \; d x_1 d x_2.
\eee
Here  the index $a$ refers to the components with respect to  Lie algebra basis $t_a$ introduced in Sect.~\ref{consec}, so 
\bee
[\partial_i n , n]_a = (t_a,[\partial_i n , n]).
\eee
In order to have a translation-invariant theory, we assume that the spiralization tensor $\mathcal D$ is constant. 

We now write  the model \eqref{Skyrme2}  as a gauged non-linear sigma model for  a translation-invariant  $SU(2)$  gauge potential 
\bee
\label{genA}
A=A_1dx_1 +A_2 dx_2,
\eee
where $A_1,A_2$ are  Lie-algebra valued constants. Noting that the curvature is 
\bee
F = [A_1,A_2] \; dx_1\wedge dx_2, 
\eee
the energy functional 
\eqref{localenergy} of the gauged non-linear sigma model
takes the form
\begin{align}
\label{constgaugefield}
E[A,n]& =\int_{\R^2} \left( \frac{1}{2} |\partial_1 n|^2  + \frac{1}{2} |\partial_2n|^2 -(A_1,[\partial_1 n, n])  - (A_2, [\partial_2n,n]) \right. \nonumber  \\
& \qquad \qquad \qquad   \left.   +  \frac 12 |[A_1,n]|^2 +  \frac 12 |[A_2,n]|^2  -(n, [A_1,A_2]) \right) d x_1  d x_2.
 \end{align} 

The key step in  the translation from  magnetic skyrmions to gauged non-linear  sigma models is the identification of the gauge field with the spiralization tensor according to 
\bee
\label{AfromD}
A_i =-\sum_{a=1}^3 {\mathcal D}_{ai} t_a, \quad i=1,2.
\eee
In other words, the spiralization tensor is interpreted as minus the matrix  obtained when expanding $A_1, A_2$ in  the basis $t_a$ of $su(2)$. This prescription  should be viewed as special case of the more natural three-dimensional situation, where we have a further component $A_3$ of the gauge field and the spiralization tensor is   minus the   $3\times 3$ matrix representing the linear map which takes  the basis elements  $t_a$ into $A_i$, $i=1, 2, 3$.

The DM term can then  be written in terms of the gauge field as 
\bee
 \sum_{a=1}^3\sum_{i=1}^2 {\mathcal D}_{ai}[\partial_i  n , n]_a =  -\sum_{i=1}^2 (A_i,[\partial_i n, n]). 
 \eee
If we  now pick the potential
\bee
\label{specpot}
V_A(n) = \frac 12 |[A_1,n]|^2 +  \frac 12 |[A_2,n]|^2  -(n, [A_1,A_2]),
\eee
then the energy functional \eqref{Skyrme2} for magnetic skyrmions with the potential $V=V_A$ equals the 
expression  \eqref{constgaugefield} with 
the particular gauge field \eqref{AfromD}. This observation is one of the key results of this paper,  and allows us to  obtain stationary points of the magnetic skyrmion energy \eqref{Skyrme2}  by   solving   the Bogomol'nyi equation \eqref{Bogonhol} for the gauged non-linear sigma model with gauge field \eqref{AfromD}. The choice of potential \eqref{specpot} for any given spiralization tensor  generalises the notion of `critically coupled' introduced in  \cite{BSRS}, and we will use this term to describe the solvable models of magnetic skyrmions defined by \eqref{Skyrme2} with $V=V_A$.  

Solving the Bogomol'nyi equations turns out to be straightforward.  Noting that 
 \bee
A_{\bar z} = \frac 12 (A_1+iA_2) =  g\partial_{\bar z} g^{-1} , 
 \eee 
for constant $A_1$ and $A_2$,  is solved by 
 \bee
g= \exp(- \frac 12 (A_1+iA_2)\bar z),
 \eee
we obtain  the general solution from    \eqref{mastersol}. Since $A_{\bar z}$ is generally a complex, traceless $2\times 2$ matrix, 
 the explicit form of $g$ as a $2\times 2 $ matrix can  be calculated by observing  that $A_{\bar z}$ is   conjugate (by a $SL(2,\C)$ matrix) either to a diagonal matrix with equal and opposite complex eigenvalues (the generic case) or to a nilpotent  matrix. We consider and interpret the special cases  where  $A_{\bar z}$ is  nilpotent or has purely imaginary eigenvalues in some detail below. 

Regarding the general case, we note that, if 
 \bee
 \label{diagA}
A_{\bar z} = \frac 12 \begin{pmatrix} \lambda + i\omega & 0 \\ 0 & -\lambda -i\omega \end{pmatrix},
 \eee 
then the general  solution  \eqref{mastersol} is 
\bee
\label{diagsol}
w= e^{(\lambda +i\omega)\bar z} f(z),
\eee
for some holomorphic map $f$. If  $A_{\bar z}$ is conjugate to \eqref{diagA} via a constant $h\in SL(2,\C)$ then the corresponding solution is obtained by acting with $h$ on \eqref{diagsol} via fractional linear transformations  according to \eqref{fractrans}; the magnetisation field $n$ is obtained via \eqref{wn} as before.

\subsection{Axisymmetric DM interactions}
The Dirichlet energy term in \eqref{Skyrme2} is invariant under translations, reflections and  rotations in the plane and  under reflections and  rotations (about any axis) of the magnetisation field $n$.  The DM interaction  breaks this symmetry, and for generic  but  constant
spiralization tensors  the breaking is maximal, leaving only the translational symmetry intact.   However, for particular choices of $\mathcal D$, the  DM term is invariant under  rotations and reflections in the plane and simultaneous rotations and reflections  of the magnetisation vector $n$. The symmetry group is isomorphic to $O(2)$ and we call such DM terms axisymmetric. They are easily characterised  in terms of our gauge potential $A$. The DM term is axisymmetric if and only if a spatial rotation of the gauge field 
\bee
\label{sparot}
A_1 \mapsto \cos\beta \, A_1 + \sin\beta \, A_2, \qquad A_2 \mapsto -\sin \beta \,  A_1 + \cos\beta \, A_2,
\eee
can be written as a  rotation in the $su(2)$ Lie algebra, i.e., by a conjugation of $A_1$ and $A_2$ with a $SU(2)$ matrix. This is the case if and only if 
\bee
\label{orthon}
|A_1|^2 = |A_2|^2, \qquad (A_1,A_2)=0,
\eee
 i.e. if $A_1, A_2$  and $[A_1,A_2]$ form  an, up to scale, orthonormal basis of $su(2)$.  In this  case, there exists  a rescaling  by $\kappa  >0 $ and a $SU(2)$ matrix    $u$  so that 
 \bee
 \label{specond}
 \kappa ut_1u^{-1}= A_1, \quad \kappa  ut_2u^{-1}= A_2, \quad  \kappa^2ut_3u^{-1}= [A_1,A_2].
\eee
 The DM term is then invariant under spatial rotations \eqref{sparot}  and the  simultaneous rotation 
 of the magnetisation vector  according to
 \bee
 n \mapsto R(u)R_3(\beta) R(u)^{-1} n,
 \eee
 where $R(u)$ is the $SO(3)$ matrix associated to the $SU(2)$ matrix $u$ via \eqref{uad} and $R_3(\beta)$ is the rotation about $t_3$ by $\beta$. 
 It is also invariant under reflections 
 \bee
A_1 \mapsto  A_1, \qquad A_2 \mapsto - A_2, \qquad  n \mapsto R(u)S_{13} R(u)^{-1} n,
\eee
where $S_{13}$ is the reflection in the $13$ plane. 
 Note also that the potential  \eqref{specpot} takes a particularly simple form when \eqref{orthon} holds:
 \bee
 V_A(n) = \frac {1}{2\kappa^2}  \left(\kappa^2 - (n,[A_1,A_2])\right)^2.
 \eee
 
 The condition \eqref{orthon} also leads to a considerable simplification in the solution of the model. It is easy to check that it is  equivalent to the matrix  $A_{\bar z}$ being nilpotent and therefore conjugate (via the  $SU(2)$ matrix $u$)  to a matrix of the form
 \bee
 \label{uptriang}
 \begin{pmatrix}
 0 & * \\ 0 & 0
 \end{pmatrix},
 \eee
 for some complex entry $*$. This is the  case considered in \cite{BSRS}, which dealt with
  the  DM term 
  \bee \sum_{a=1}^3\sum_{i=1}^2 {\mathcal D}_{ai}[\partial_i n , n]_a = 
  \kappa \cos\alpha \; w_B +\kappa \sin\alpha  \; w_N,
  \eee
where 
\begin{align}
w_B& = \phantom{-} n_1\partial_2n_3 -n_2\partial_1 n_3 + n_3(\partial_1n_2 -\partial_2n_1),\nonumber \\ 
w_N& = - n_1\partial_1n_3 +n_2\partial_2 n_3 + n_3(\partial_1n_1 +\partial_2n_2).
\end{align} 
 In  our notation this corresponds to the spiralization tensor
 \bee
 \mathcal{D} = \kappa \begin{pmatrix} \phantom{-}\cos\alpha  & \sin \alpha \\ -\sin \alpha & \cos \alpha \\ 0 & 0 \end{pmatrix},
 \eee
 and hence 
 \bee
 A_1= -(\cos \alpha \; t_1 - \sin \alpha  \; t_2) \quad A_2=-(\sin \alpha  \;t_1  +\cos\alpha \;  t_2),
 \eee
or, with the conventions \eqref{tpm}, 
 \bee
\label{Askyr}
A_{\bar z} = 
 -\frac{1}{2}\kappa e^{i\alpha}  t_+ = \begin{pmatrix}  0 &  \frac{i}{2}\kappa e^{i\alpha} \\ 0 & 0 \end{pmatrix}.
\eee
In this case the  potential is simply
\bee
V_A(n)=\frac {\kappa^2}{ 2 } (1-n_3)^2.
\eee

 To apply our method of solution, we note that 
\bee
\label{rhoskyr}
g = \exp\left(\frac{\kappa}{2} e^{i\alpha} \bar z t_+\right) =\begin{pmatrix}
1 & - \frac{ i }{2} \kappa  e^{i\alpha} \bar z \\ 0 & 1
\end{pmatrix}
\eee
solves \eqref{Agrel} for the gauge field \eqref{Askyr}. Thus, the general solution  \eqref{mastersol} is  given in terms of a holomorphic function $f:\C \rightarrow \CP^1$  and the fractional linear transformation \eqref{fractrans} as 
\bee
\label{standskyr}
w= g[f]= \frac{f}{1 - \frac{ i }{2} \kappa e^{i\alpha} \bar z f}. 
\eee
In terms of $v=1/w$ this is 
\bee
v= - \frac{ i }{2} \kappa e^{i\alpha} \bar z + \frac{1}{f},
\eee
which, after  re-naming $f\rightarrow 1/f$, is the general solution found  and discussed in \cite{BSRS}.

In the more general case \eqref{orthon}, the  solutions of the model are 
obtained from solutions \eqref{standskyr} by rotating with $R(u)$. However,  the physics is quite different.  Even if  both $A_1$ and $A_2$ are in the $t_1 t_2$-plane,  the two possible directions of $[A_1,A_2]$    lead to different chiralities \cite{genDMI,BSRS}.   In the generic case,  we obtain a linear combination of `in plane' and `out of plane' DM terms.  This illustrates that a rotation, which is a   change of  gauge in the gauged sigma model, can  make a physical difference in the interpretation as a model  of magnetic skyrmions.

\subsection{Rank one  DM interaction}  
 The  spiralization matrix $\mathcal D$, still assumed to be constant,  has rank 1 when $A_1$ and $A_2$ are constant and collinear, so when 
 \bee
  [A_1,A_2]=0. 
 \eee
 In this case,   the curvature $F$ of $A$ vanishes. As a result, the gauge field can be removed entirely by a gauge transformation. The solutions are then  related to the Belavin-Polyakov solitons of the $O(3)$ sigma model \cite{BP} (holomorphic or anti-holomorphic maps $\Sigma\rightarrow \CP^1$) by space-dependent $SU(2)$ transformations. Our general formula \eqref{mastersol} reproduces the solutions related to holomorphic maps. For example 
 \bee
 \label{rank1}
A_1= a t_3, \qquad A_2=A_3 =0, \qquad a\in \R,
\eee
leads to the potential
 \bee
V_A(n) = \frac {a^2} {2}  (n_1^2 +n_2^2) = \frac {a^2} {2}  (1-n_3^2), 
 \eee
and  the general solution 
\bee
w =  e^{-\frac{ i}{2}   a \bar z} f(z),
\eee
where $f$ an arbitrary holomorphic function. The simplest case is $f=1$, and produces the  configuration
\bee
n = \begin{pmatrix} \phantom{-} \sech \left( \frac a 2 x_2 \right)  \cos \left(\frac a 2 x_1\right)  \\
- \sech \left( \frac a 2 x_2 \right) \sin\left(\frac a 2 x_1\right) \\
\tanh\left(\frac a 2 x_2\right)
\end{pmatrix} .
\eee
This is a kink interpolating between  the `up' and the `down'  vacuum of the potential  in the $x_2$-direction and rotating in the $x_1$-direction at the same time. One can get rid of the kink (leading to a helix) or of the rotation (leading to a pure kink) by choosing 
\bee
f(z) = e^{-i\frac  a2  z}\quad \text{or} \quad  f(z) = e^{i\frac  a2  z}.
\eee
If one picks 
\bee
f(z) = e^{-i\frac  a2  z}r_n(z) ,
\eee
where $r_n$ is a rational map of degree $n$, one obtains
\bee
\label{hellump}
w=  e^{-i  a  x_1}r_n (z),
\eee
i.e. a  rational  map modulated by a helix in the $x_1$-direction.  This describes a  Belavin-Polyakov  multi-soliton \cite{BP}  modulated by a helix. 

\subsection{Impurities as non-abelian gauge fields}
\label{impuresect}
To end our discussion of applications,  we indicate how  the interaction of Belavin-Polyakov solitons with impurities can also be described in terms of the  gauged non-linear sigma model introduced in this paper.  Unlike in the application to magnetic skyrmions, the non-abelian connection will need a non-trivial spatial variation to model an impurity.  

In  the recent paper  \cite{AQW}, the  authors study  mathematical models  for Belavin-Polyakov solitons which interact with impurities in a way which  preserves supersymmetry.  The  first order equation for solitons in the presence of an impurity proposed in that paper 
is rather similar to \eqref{Bogonhol}, but with a given impurity configuration instead of a gauge field. We will now   derive the impurity equation of \cite{AQW} from our Bogomol'nyi equation \eqref{Bogonhol}, explain  how to pick the gauge field to reproduce the impurities studied in\cite{AQW} and point out generalisations.

Extending the discussion of \cite{AQW} to an arbitrary Riemann surface, and using the complex notation introduced in Sect.~\ref{stereo} to parametrise $\CP^1$ in terms of $\C\cup \{\infty\}$,   the impurity field considered in\cite{AQW} is a map 
\bee
\sigma: \Sigma \rightarrow \CP^1.
\eee
Further adapting conventions so that the soliton field $u$ considered in \cite{AQW} is our field $\bar w$, 
the simplest equation of \cite{AQW}   for a configuration $w$  coupled to such an impurity  is
\bee
\label{impure}
\partial_{\bar z} w + \sigma =0. 
\eee
To see that this is a special case of our Bogomol'nyi equation \eqref{Bogonhol}, we use the expansion \eqref{complexcov} to write \eqref{Bogonhol} as 
\bee
\label{spelledout}
\partial_{\bar z} w - \frac{i}{2} (A_+)_{\bar z} + i(A_3)_{\bar z} w + \frac{i}{2} (A_-)_{\bar z} w^2 = 0,
\eee
where we remind the reader that the indices $+,-, 3$ refer to the Lie algebra components \eqref{Aexp}. To obtain \eqref{impure}  we simply need to pick
\bee
- \frac{i}{2} (A_+)_{\bar z}  =\sigma,  \quad (A_3)_{\bar z}= \frac{i}{2} (A_-)_{\bar z} =0,
\eee
which amounts to
\bee
A_{\bar z}  = \begin{pmatrix}  0 & 0 \\ \sigma & 0 \end{pmatrix}, \qquad 
A_{z}  = \begin{pmatrix}  0 & -\bar {\sigma}  \\ 0  & \phantom{-}0 \end{pmatrix}.
\eee
One checks that this connection   can be expressed  according to 
\eqref{Agrel}  in terms of the $SL(2,\C)$-valued map 
\bee
g =\begin{pmatrix}
1 & 0 \\ - \int \sigma d \bar z  & 1 
\end{pmatrix}, 
\eee
so that our  general solution  \eqref{mastersol}  in this case is  
\bee
\label{genimp}
w =  f - {\textstyle \int} \sigma d \bar z,
\eee
with  $f$ again holomorphic. This is the obviously the general solution of \eqref{impure}, and was also  derived and studied in \cite{AQW}. However, as also discussed in \cite{AQW},  simple properties like total energy and degree of configurations like \eqref{genimp} are rather subtle, despite the explicit form, echoing a similar observation regarding exact magnetic skyrmions  in \cite{BSRS}. 

The authors of \cite{AQW} also   discuss Bogomol'nyi equations for solitons interacting with  impurities which contain  products of the impurity field and the  soliton field. In our general equation \eqref{spelledout} this would simply amount to picking a connection $A$ with a non-trivial component $A_3$. Our general solution again applies.  The possibility of interactions which are quadratic in the soliton field correspond to a non-trivial component $A_-$ of the gauge field.  This  does not appear to have been considered in the literature on impurities, but would seem equally natural from our point of view.

 \section{Conclusion} 

In this paper we introduced and studied the  gauged  non-linear sigma model defined by the energy functional \eqref{energy}. We showed that it  provides a framework for systematically studying solvable theories of magnetic skyrmions with  any given DM interaction term. We also indicated  that it provides a natural language for  solvable  theories of Belavin-Polyakov solitons interacting with impurities.

We showed  how to  solve the  first order  Bogomol'nyi equation \eqref{Bogonhol} of the gauged non-linear sigma model by exploiting the relation between holomorphic and unitary gauges. This leads to the explicit formula \eqref{mastersol} for  local solutions. In the application  considered here we assumed $\Sigma =\C$, in which case we obtain  infinitely many globally defined magnetic skyrmions  for any given  spiralization tensor, and similarly infinitely many globally defined solitons in the presence of any given impurity. 

Summed up as a slogan, our results show that the geometry behind both  DM interactions and impurities in the $O(3)$ sigma model is a Chern connection.  They also suggest that DM interactions and impurities  may be  different aspects of  the same underlying physics, and that tools used in the study of one may be usefully applied to the other. For example, the supersymmetric nature of the model for impurities studied in \cite{AQW} suggests that the non-abelian sigma model studied here also has a natural supersymmetric extension. This may, in turn, have interesting implications for magnetic skyrmions in the theories we studied here. 

From a mathematical,   and possibly  also physical,   point of view it  would certainly be of interest to  repeat our analysis on more general  Riemann surfaces, and to study global properties of solutions  \eqref{mastersol} there. This requires a choice of unitary connection.  As we explained, such  connections define complex structures on $\C^2$-bundles and hence on associated $\CP^1$-bundles over such Riemann surfaces, and this  provides  a natural geometrical interpretation for any chosen connection.

It would also be interesting to consider Riemann surfaces with boundary, and to clarify the correct boundary conditions  in each case. We only briefly touched on the relevant issues in Sect.~\ref{boundarysect} of this paper.

\vspace{0.2cm}

\section*{Acknowledgements} Some of the results in this paper were  presented  in  two seminars I gave  in Edinburgh in early 2019. I thank  members of the audience for insightful comments,   Lorenzo Foscolo for several discussions and Calum Ross for comments on an earlier version of this manuscript. I also thank the SciPost referees for constructive comments which helped   improve the presentation. 

\section*{Note added in proof}  During the proof stage of this paper,  the e-print \cite{Walton} appeared on the arxiv in which the author points out that the gauged sigma model and the Bogomol'nyi equation proposed here can naturally be expressed and generalised in the language of $J$-holomorphic curves and equivariant cohomology, as discussed in \cite{CGS}

\end{document}